\documentclass[useAMS,usenatbib,usegraphicx,usebibtex8]{mn2e}

\title[A Bayesian Re-analysis of the Gliese 667C HARPS Data]{Additional Keplerian Signals in the HARPS data for Gliese 667C from a Bayesian Re-analysis}
\author[P. C. Gregory]{Philip C. Gregory$^{1}$\thanks{E-mail:
gregory@phas.ubc.ca} \\
$^{1}$Physics and Astronomy Department, University of British Columbia, 6224 Agricultural Rd., Vancouver, BC V6T 1Z1, Canada}

\begin{document}

\date{Submitted to MNRAS Dec. 16, 2012 }

\pagerange{\pageref{firstpage}--\pageref{lastpage}} \pubyear{2002}

\maketitle

\label{firstpage}

\begin{abstract}
A re-analysis of Gliese 667C HARPS precision radial velocity data was carried out with a Bayesian multi-planet Kepler periodogram (from 0 to 7 planets) based on a fusion Markov chain Monte Carlo algorithm. The most probable number of signals detected is 6 with a Bayesian false alarm probability of 0.012. The residuals are shown to be consistent with white noise. The 6 signals detected include two previously reported with periods of 7.198 (b) and 28.14 (c) days, plus additional periods of 30.82 (d), 38.82 (e),  53.22, and  91.3 (f) days. The 53 day signal is probably the second harmonic of the stellar rotation period and is likely the result of surface activity. The existence of the additonal Keplerian signals suggest the possibilty of further planets, two of which (d and e)  could join Gl 667Cc in the central region of the habitable zone. N-body simulations are required to determine which of these signals are consistent with a stable planetary system. $M \sin i$ values corresponding to signals b, c, d, e, and f are $\sim$ 5.4, 4.8, 3.1, 2.4, and 5.4 M$_{\earth}$, respectively. 
 
\end{abstract}


\section{Introduction}

The HARPS spectrograph mounted on the ESO/3.6-m La Silla telescope is yielding a rich bounty of information regarding planets around M dwarfs (eg., \citealt{Bonfils2011}). There is a lot of interest in searches for planets about low mass M dwarfs. Firstly, a planet of given mass and orbital separation induces a larger stellar radial velocity variation around a lower mass star. Secondly, the low luminosity of M dwarfs moves their habitable zone much nearer to the star. For these reasons a habitable zone (HZ) planet around a 0.3 M$_{\sun}$ M dwarf produces a 7 times larger radial velocity wobble than the same planet orbiting
a solar-mass G dwarf \citep{Delfosse2012}.

One M dwarf of particular interest is Gliese 667C (Gl 667C, also known as GJ 667C), an isolated component of a hierarchical triple system. The two others components, Gl 667AB, are a closer couple of K dwarfs. Gl 667AB has a semi-major axis of 1.82 AU (period of 42.15 years) and a total mass of 1.27 M$_{\sun}$ (dynamically determined, \citealt{Soderhjelm1999}). Gl 667C is the lightest component with a mass of $0.310\pm 0.019$  M$_{\sun}$  \citep{Anglada2012a}. It is at a projected distance of 32.4'' from Gl 667AB, giving an expected semi-major axis of $\sim 300$ AU (for a distance of 7.23 pc and a factor of 1.26 between expected and projected semi-major axis, \citealt{Fischer1992}).

In 2011, \cite{Bonfils2011} reported a super-earth Gl 667Cb ($M \sin i = 5.9$ M$_{\earth}$) with a period of 7.2 d and evidence for two other interesting periods of 90 and 28 d. The possibility of a 28 d period planet was particularly interesting because it would fall in the HZ. Two recent papers (\citealt{Anglada2012a} \& \citealt{Delfosse2012}), have confirmed planet Gl 667Cb and a 28 d period planet Gl 667Cc ($M \sin i = 4.3$ M$_{\earth}$) in the HZ. The \cite{Anglada2012a} results are not fully independent as they depend on the 143 \cite{Bonfils2011} RV observations which they reduced using the HARPS-TERRA (Template Enhanced Radial velocity Re-analysis Application) algorithm \citep{Anglada2012b}, supported by observations from two other spectrographs. The \cite{Delfosse2012} data includes an additional 36 HARPS radial velocity measurements.

The excitement generated by this and many other exoplanetary discoveries has spurred a significant effort to improve the statistical tools for analyzing data in this field (e.g., \citealt{LoredoChernoff2003}, \citealt{Loredo2004}, \citealt{Cumming2004}, Gregory 2005a \& b, Ford 2005 \& 2006, \citealt{FordGregory2007}, \citealt{Tuomi2009}, \citealt{Dawson2010}, \citealt{CummingDragomir2010}). Much of this work has highlighted a Bayesian MCMC approach as a way to better understand parameter uncertainties and degeneracies and to compute model probabilities. 

Gregory (2011, 2011b, 2012) developed a Bayesian fusion MCMC algorithm that incorporates parallel tempering (PT), simulated annealing and a genetic crossover operation to facilitate the detection of a global minimum in $\chi^2$. This enables the efficient exploration of a large model parameter space starting from a random location. When implemented with a multi-planet Kepler model~\footnote{For multiple planet models, there is no analytic expression for the exact radial velocity perturbation. In many cases, the radial velocity perturbation can be well modeled as the sum of multiple independent Keplerian orbits which is what has been assumed in this paper.}, it is able to identify any significant periodic signal component in the data that satisfies Kepler's laws and is able to function as a multi-planet Kepler periodogram~\footnote{Following on from the pioneering work on Bayesian periodograms by \citet{Jaynes1987} and \citet{Brett1988}}. The fusion MCMC algorithm includes an innovative adaptive control system that automates the selection of efficient parameter proposal distributions even if the parameters are highly correlated. One application of the algorithm (\citealt{Gregory2010}) confirmed the existence of a disputed second planet (\citealt{Fischer2002}) in 47 Ursae Majoris (47 UMa) and provided orbital constraints on a possible additional long period planet with a period $\sim 10000$d.

This paper presents a Bayesian re-analysis of the HARPS data \citep{Delfosse2012} for the star Gl 667C using our fusion MCMC based multi-planet Kepler periodogram. Section~\ref{sec:fusion} provides a brief overview of our Bayesian approach and outlines the adaptive fusion MCMC algorithm (\citealt{Gregory2011b} and \citealt{Gregory2012}). Section~\ref{sec:analysis} gives the model equations and priors. Sections~\ref{sec:HARPS} and \ref{sec:modsel} present the parameter estimation and model selection results. The final two sections are devoted to discussion and conclusions.

\section{The adaptive fusion MCMC}
\label{sec:fusion}

The adaptive fusion MCMC (FMCMC) is a very general Bayesian nonlinear model fitting program. After specifying the model, $M_i$, the data, $D$, and priors, $I$, Bayes theorem dictates the target joint probability distribution for the model parameters which is given by
\begin{equation}
p(\vec{X}|D,M_{i},I) = C \ p(\vec{X}|M_{i},I) \times p(D|M_{i},\vec{X},I).
\label{eq:orbit25}
\end{equation} 
where $C$ is the normalization constant and $\vec{X}$ represent the set of model parameters. The first term on the RHS of the equation, $p(\vec{X}|M_i,I)$, is the prior probability distribution of $\vec{X}$, prior to the consideration of the current data $D$. The second term, $p(D|\vec{X},M_i,I)$, is called the likelihood and it is the probability that we would have obtained the measured data $D$ for this particular choice of parameter vector $\vec{X}$, model $M_i$, and prior information $I$. At the very least, the prior information, $I$, must specify the class of alternative models (hypotheses) being considered (hypothesis space of interest) and the relationship between the models and the data (how to compute the likelihood). In some simple cases the log of the likelihood is simply proportional to the familiar $\chi^2$ statistic. For further details of the likelihood function for this type of problem see \cite{Gregory2005b}.

To compute the marginals for any subset of the parameters it is necessary to integrate the joint probability distribution over the remaining parameters. For example, the marginal probability distribution (a probability density function) of the orbital period in a one planet radial velocity model fit is given by
\\

\begin{eqnarray}
p(P|D,M_{1},I) & = & \int dK \int dV \int de \int d\chi \nonumber\\
& &  \times \;  \int d\omega  \int dslope\int ds \nonumber\\
& &  \times \;  p(P,K,V,e,\chi,\omega,slope,s|D,M_{1},I) \nonumber\\
& \propto & p(P|M_{1},I) \int dK \cdots \int ds \nonumber\\
& & \times \; p(K,V,e,\chi,\omega,slope,s|M_{1},I) \nonumber\\ 
& & \times \; p(D|M_{1},P,K,V,e,\chi,\omega,slope,s,I),\nonumber\\ 
\label{eq:orbit}
\end{eqnarray} 
where $P,K,V,e,\chi,\omega$ are the radial velocity model parameters and $s$ is an extra noise parameter which is discussed in Section~\ref{sec:analysis}. $slope$ is a parameter that accounts for the long period orbital motion induced in Gl 667C by Gl 667AB. $p(P|M_{1},I)$ is the prior for the orbital period parameter, $p(K,V,e,\chi,\omega,slope,s|M_{1},I)$ is the joint prior for the other parameters, and $p(D|M_{1},P,K,V,e,\chi,\omega,slope,s,I)$ is the likelihood. For a seven planet model fit we need to integrate over 37 parameters to obtain  the marginal probability distribution for each of the seven period parameters. Integration is more difficult than maximization, however, the Bayesian solution provides the most accurate information about the parameter errors and correlations without the need for any additional  calculations, i.e., Monte Carlo simulations. Bayesian model selection requires integrating over all the model parameters.

In high dimensions, the principle tool for carrying out the integrals is Markov chain Monte Carlo based on the Metropolis algorithm. The greater efficiency of an MCMC stems from its ability, after an initial burn-in period, to generate  samples in parameter space in direct proportion to the joint target probability distribution. In contrast, straight Monte Carlo integration randomly samples the parameter space and wastes most of its time sampling regions of very low probability. 

MCMC algorithms avoid the requirement for completely independent samples, by constructing a kind of random walk in the model parameter space such that the number of samples in a particular region of this space is proportional to a target posterior density for that region. The random walk is accomplished using a Markov chain, whereby the new sample, $\vec{X}_{t+1}$, depends on previous sample $\vec{X}_t$ according to a time independent entity called the {\it transition kernel}, $p(\vec{X}_{t+1}|\vec{X}_t)$. The remarkable property of $p(\vec{X}_{t+1}|\vec{X}_t)$ is that after an initial \index{Burn-in period}burn-in period (which is discarded) it generates samples of $\vec{X}$ with a probability density proportional to the desired posterior $p(\vec{X}|D,M_{1},I)$ (e.g., see Chapter 12 of \cite{Gregorybook} for details). 

The joint posterior probability distribution in model parameter space is typically highly multi-modal for exoplanet radial velocity (RV) analysis. The Metropolis algorithm can become stuck in the vicinity of a local probability maximum. To avoid this fusion MCMC (FMCMC) incorporates three other algorithms each of which is designed to facilitate the detection of a global minimum in  $\chi^2$ (or a maximum in probability). They are parallel tempering (PT)  (\citealt{Geyer1991} and re-invented by \citealt{Hukushima1996}), simulated annealing and a genetic crossover operation. All three are implement in each FMCMC run. The combination greatly facilitates the identification of a global maximum in probability. This was made possible through the development of an adaptive control system which has been described in detail most recently by Gregory (2011b, 2012). Further refinements of the control system are ongoing.

At each iteration of the FMCMC, a single joint proposal is made to jump to a new location in parameter space from the current location. The key to an efficient MCMC is an efficient method of proposing new jumps especially when there are correlations between the parameters. This is further complicated in PT because multiple MCMC are executed in parallel each exploring a different probability distribution. In FMCMC this difficult step is automated by the control system saving a great deal of time and effort.

\section{Models and Priors}
\label{sec:analysis}

In this section we describe the model fitting equations and the selection of priors for the model parameters. We have investigated the Gl 667C data using models ranging from 0 to 7 planets. For a one planet model the predicted radial velocity is given by
\begin{equation}
v(t_i) = V + K [\cos\{\theta(t_i+\chi P)+\omega\} + e \cos \omega],
\label{eq:orbit1}
\end{equation}
and involves the 6 unknown parameters
\begin{itemize}
\item[] $V =$ a constant velocity.
\item[] $K =$ velocity semi-amplitude. 
\item[] $P =$ the orbital period.
\item[] $e =$ the orbital eccentricity.
\item[] $\omega =$ the longitude of periastron.
\item[] $\chi =$ the fraction of an orbit, prior to the start of data taking, that periastron occurred at. Thus, $\chi P =$ the number of days prior to $t_i = 0$ that the star was at periastron, for an orbital period of P days. 
\item[] $\theta(t_i+\chi P) =$ the true anomaly, the angle of the star in its orbit relative to periastron at time $t_i$.
\end{itemize}

We utilize this form of the equation because we obtain the dependence of $\theta$ on $t_i$ by solving the conservation of angular momentum equation
\begin{equation}
\frac{d\theta}{dt} - \frac{2 \pi [1+e\cos \theta(t_i+\chi \; P)]^2}{P (1-e^2)^{3/2}} = 0.
\label{eq:orbit2}
\end{equation}
Our algorithm is implemented in {\it Mathematica} and it proves faster for {\it Mathematica} to solve this differential equation than solve the equations relating the true anomaly to the mean anomaly via the eccentric anomaly. {\it Mathematica} generates an accurate interpolating function between $t$ and $\theta$ so the differential equation does not need to be solved separately for each $t_i$. Evaluating the interpolating function for each $t_i$ is very fast compared to solving the differential equation. Details on how equation~\ref{eq:orbit2} is implemented and the accuracy of the interpolation as a function of eccentricity are given in the Appendix A of \citet{Gregory2011b}.

As described in more detail in \citealt{Gregory2007}, we employed a re-parameterization of $\chi$ and $\omega$ to improve the MCMC convergence speed motivated by the work of \cite{Ford2006}. The two new parameters are $\psi=2\pi\chi+\omega$ and $\phi=2 \pi\chi-\omega$. Parameter $\psi$ is well determined for all eccentricities. Although $\phi$ is not well determined for low eccentricities, it is at least orthogonal to the $\psi$ parameter. We use a uniform prior for $\psi$ in the interval 0 to $4 \pi$ and uniform prior for $\phi$ in the interval $-2 \pi$ to $+2 \pi$. This insures that a prior that is wraparound continuous in $(\chi,\omega)$ maps into a wraparound continuous distribution in $(\psi,\phi)$. To account for the Jacobian of this re-parameterization it is necessary to multiply the Bayesian integrals by a factor of $(4 \pi)^{-nplan}$, where $nplan =$ the number of planets in the model. In this work we utilized the orthogonal combination $(\psi,\phi)$ as well as the FMCMC correlated proposal scheme described in \cite{Gregory2012}.

In a Bayesian analysis we need to specify a suitable prior for each parameter. These are tabulated in Table~\ref{tab:priors}. For the current problem, the prior given in equation~\ref{eq:orbit} is the product of the individual parameter priors. Detailed arguments for the choice of each prior were given in \cite{Gregory2007} and \cite{Gregory2010}.     
\begin{table*}
 \centering
 \begin{minipage}{140mm}
  \caption{Prior parameter probability distributions.}
  \label{tab:priors}
  \begin{tabular}{@{}llll@{}}
 \hline
   Parameter    &    Prior        & Lower bound & Upper bound\\
 \hline
Orbital frequency  & $p(\ln f_1, \ln f_2, \cdots \ln f_n|M_n,I) = \frac{n!}{[\ln (f_H/f_L)]^n}$  & 1/0.5 d & 1/95 yr~\footnote{ Gl 667C is part of a triple star system with Gl 667AB, a much closer binary. We adopted at an upper limit of 95 yr by setting the gravitational pull on the planet due to Gl 667AB $= 1\%$ of the pull from Gl 667C. }  \\
&\ ($n=$number of planets)  & &  \\
& & & \\
Velocity $K_i$  &  Modified scale invariant~\footnote{Since the prior lower limits for $K$ and $s$ include zero, we used a modified scale invariant prior of the form
\begin{equation}
p(X|M,I) = \frac{1}{X+X_0}\; \frac{1}{\ln\left(1+\frac{X_{\rm max}}{X_0}\right)}
\label{eq:orbit13}
\end{equation}
For $X \ll X_0$, $p(X|M,I)$ behaves like a uniform prior and for $X \gg X_0$ it behaves like a scale invariant prior. The $\ln\left(1+\frac{ X_{\rm max}}{X_0}\right)$ term in the denominator ensures that the prior is normalized in the interval 0 to $X_{\rm max}$.} & 0 \ (K$_0 = 1)$ &  $K_{\rm max}\ \left(\frac{P_{\rm min}}{P_i}\right)^{1/3} \frac{1}{\sqrt{1-e_i^2}}$ \\
\ \ \  (m s$^{-1}$) & & & \\
  & \ \ \ \ \ $\frac{(K+K_0)^{-1}}{\ln{\left[1+\frac{K_{\rm max}}{K_0} \ \left(\frac{P_{\rm min}}{P_i}\right)^{1/3} \frac{1}{\sqrt{1-e_i^2}}\right]}}$
 &  & $K_{\rm max}=2129$\\
 & & & \\
V  (m s$^{-1}$) & Uniform & $-K_{\rm max}$ & $K_{\rm max}$  \\
& & & \\
$e_i$ Eccentricity & Ecc. noise bias correction filter& 0 & 0.99 \\
& & & \\
$\chi$ orbit fraction & Uniform & $0$ & 1 \\
& & & \\
$\omega_i$ Longitude of & Uniform & $0$ & $2 \pi$ \\
\ \ \ \ periastron &  &  & \\
& & & \\
$slope$ & Uniform~\footnote{Since this parameter is common to all models including the 0 planet model, the exact upper and lower bounds are not critical for model selection. The range simply needs to be large enough so as not to effect the parameter estimates.}& $-6$ & $6$ \\
\ \ \ \ m/s/yr &  &  & \\
& & & \\
$s$ Extra noise   (m s$^{-1}$) & $\frac{(s+s_0)^{-1}}{\ln{\left(1+\frac{s_{\rm max}}{s_{0}}\right)}}$ & 0  \ (s$_0 = 1$)& $K_{\rm max}$  \\
\hline
\end{tabular}
\end{minipage}
\end{table*}

All of the models considered in this paper incorporate an extra noise parameter, $s$, that can allow for any additional noise beyond the known measurement uncertainties. We adopt an independent Gaussian distribution with a variance $s^2$ where is becomes another parameter in the model fit. Thus, the combination of the known errors and extra noise has a Gaussian distribution with variance $= \sigma_i^2 + s^2$, where $\sigma_i$ is the known measurement uncertainty for i$^{\mbox{\tiny th}}$ data point. In general, nature is more complicated than our model and known measurement errors. Marginalizing $s$ has the desirable effect of treating anything in the data that can't be explained by the model and known measurement errors as noise, leading to conservative estimates of orbital parameters~\footnote{There is an additional benefit for incorporating the extra noise parameter $s$. When the $\chi^2$ of the fit is very large, the Bayesian Markov chain automatically inflates $s$ to include anything in the data that cannot be accounted for by the model with the current set of parameters and the known measurement errors. This results in a smoothing out of the detailed structure in the $\chi^2$ surface and allows the Markov chain to explore the large scale structure in parameter space more quickly. This is similar to simulated annealing, but does not require choosing a cooling scheme. The chain begins to decrease the value of the extra noise as it settles in near the best-fit parameters. This is demonstrated in Fig. 2 of \citet{Gregory2011b} and in Section 2.1 of \citet{Gregory2012}.}. See Sections 9.2.3 and 9.2.4 of \citet{Gregorybook} for a tutorial demonstration of this point. If there is no extra noise then the posterior probability distribution for $s$ will peak at $s = 0$. The upper limit on $s$ was set equal to $K_{\rm max}$. We employed a modified scale invariant prior for $s$ with a knee, $s_0 = 1$ m s$^{-1}$. 

\subsection{Eccentricity bias}
\label{sec:eccBias}

In \cite{Gregory2010}, the velocities of model fit residuals were randomized in multiple trials and processed using a one planet version of the FMCMC Kepler periodogram. In this situation periodogram probability peaks are purely the result of the effective noise. The orbits corresponding to these noise induced periodogram peaks exhibited a well defined statistical bias~\footnote{The bias found using multiple sets of randomized residuals from a 5 planet fit to 55 Cancri combined Lick and Keck data agreed closely with the bias found for multiple sets of randomized residuals from both 2 and 3 planet fits to 47 UMa Lick data.} towards high eccentricity. We offered the following explanation for this effect. To mimic a circular velocity orbit the noise points need to be correlated over a larger fraction of the orbit than they do to mimic a highly eccentric orbit. For this reason it is more likely that noise will give rise to spurious highly eccentric orbits than low eccentricity orbits. 

\cite{Gregory2010} characterized this eccentricity bias and designed a correction filter that can be used as an alternate prior for eccentricity to enhance the detection of planetary orbits of low or moderate eccentricity. On the basis of our understanding of the mechanism underlying the eccentricity bias, we expect the eccentricity prior filter to be generally applicable to searches for low amplitude orbital signals in precision radial velocity data sets.
The probability density function for this filter is given by
\begin{eqnarray}
pdf(e) & = & 1.3889-1.5212 e^2 +0.53944 e^3\nonumber\\
& &  -1.6605(e-0.24821)^8.
\label{eq:EccDeBias}
\end{eqnarray}
Fig. 11 of \cite{Gregory2011b} demonstrates that the effect of this noise eccentricity bias correction filter on the final marginal eccentricity distributions for the low and moderate eccentricity orbits of Gl 581b, c, \& d is insignificant.

In a related study, \citet{Shen2008} explored least-$\chi^2$ Keplerian fits to synthetic radial velocity data sets. They found that the best fit eccentricities for low signal-to-noise ratio $K/\sigma \le 3$ and moderate number of observations ${\rm N_{obs}}\le 60$, were systematically biased to higher values, leading to a suppression of the number of nearly circular orbits. More recently, \citet{Zakamska2011} found that eccentricities of planets on nearly circular orbits are preferentially overestimated, with typical bias of one to two times the median eccentricity uncertainty in a survey, e.g., 0.04 in the Butler et al. catalogue \citep{Butler2006}. When performing population analysis, they recommend using the mode of the marginalized posterior eccentricity distribution to minimize potential biases.

In the analysis of the Gl 667C data we used the eccentricity noise bias correction filter as the eccentricity prior on fits of all the models.

\section{Analysis of the HARPS data}
\label{sec:HARPS}

\begin{figure}
\includegraphics[width=85mm]{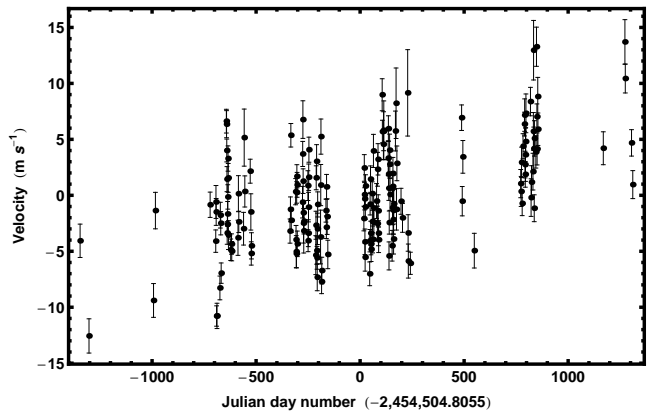}
\caption{The HARPS data for Gl 667C after correcting for the secular acceleration.}
\label{fig:HARPSdata}
\end{figure}

For all the analysis we used the HARPS data given by \citet{Delfosse2012}. We subtracted a mean velocity of 6.5474477 km s$^{-1}$ and converted to units of m s$^{-1}$. Following \citet{Delfosse2012} we also subtracted a component due to the secular acceleration of 0.21 m s$^{-1}$ yr$^{-1}$ arising from the proper motion. This small correction resulted in a change in radial velocity over the span of the data amounting to 1.527 m s$^{-1}$. Fig.~\ref{fig:HARPSdata} shows the corrected HARPS data for Gl 667C which exhibits a large positive slope indicative of a period much longer than the data.  A likely explanation is the orbital motion of Gl 667C relative to the center of mass of the Gl 667ABC system. The Gl 667ABC system parameters suggested an orbital period of $\sim 3900$ yr. For a circular orbit this implied a maximum expected K value for Gl 667C of $\sim 1850$ m s$^{-1}$ and a maximum rate of change in radial velocity of $\sim 3 $ m s$^{-1}$ yr$^{-1}$. Multiplying by the data duration of 7.27 yr yields a maximum expected velocity change of 20 m s$^{-1}$, which is comparable to what is observed. Our zero reference time is the mean time of the HARPS observations which corresponds to a Julian day number $= 2,454,504.8055$. Following \citet{Delfosse2012}, we deal with the long period velocity change by including a linear drift which we refer to as a $slope$ parameter in all our models.

The models considered range from a zero planet model to a seven planet model. As mentioned all models include a $slope$ parameter and extra noise $(s)$ parameter. The current section deals with model parameter estimation while Section~\ref{sec:modsel} deals with model selection. The next 6 subsections show the FMCMC based Kepler periodograms for the 2 - 7 planet cases. We don't show Kepler periodogram for the one planet model as its parameters are well understood.   

\subsection{Two planet model}
\label{sec:2HARPS}

\begin{figure}
\includegraphics[width=85mm]{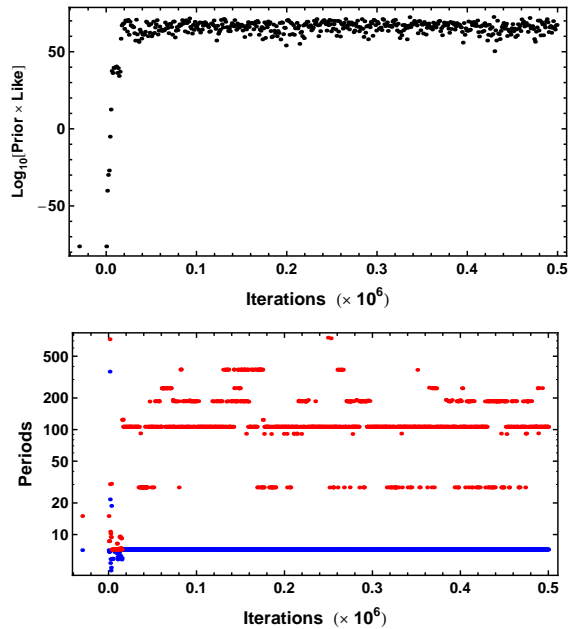}
\caption{The upper panel is a plot of the Log$_{10}$[Prior $\times$ Likelihood] versus iteration for a 2 planet Kepler periodogram of the HARPS data. The lower shows the values of the two unknown period parameters versus iteration number. The two starting periods of 7.2 and 15 d are shown on the left hand side of the plot at a negative iteration number.}
\label{fig:2planPiter}
\end{figure}

\begin{figure}
\includegraphics[width=85mm]{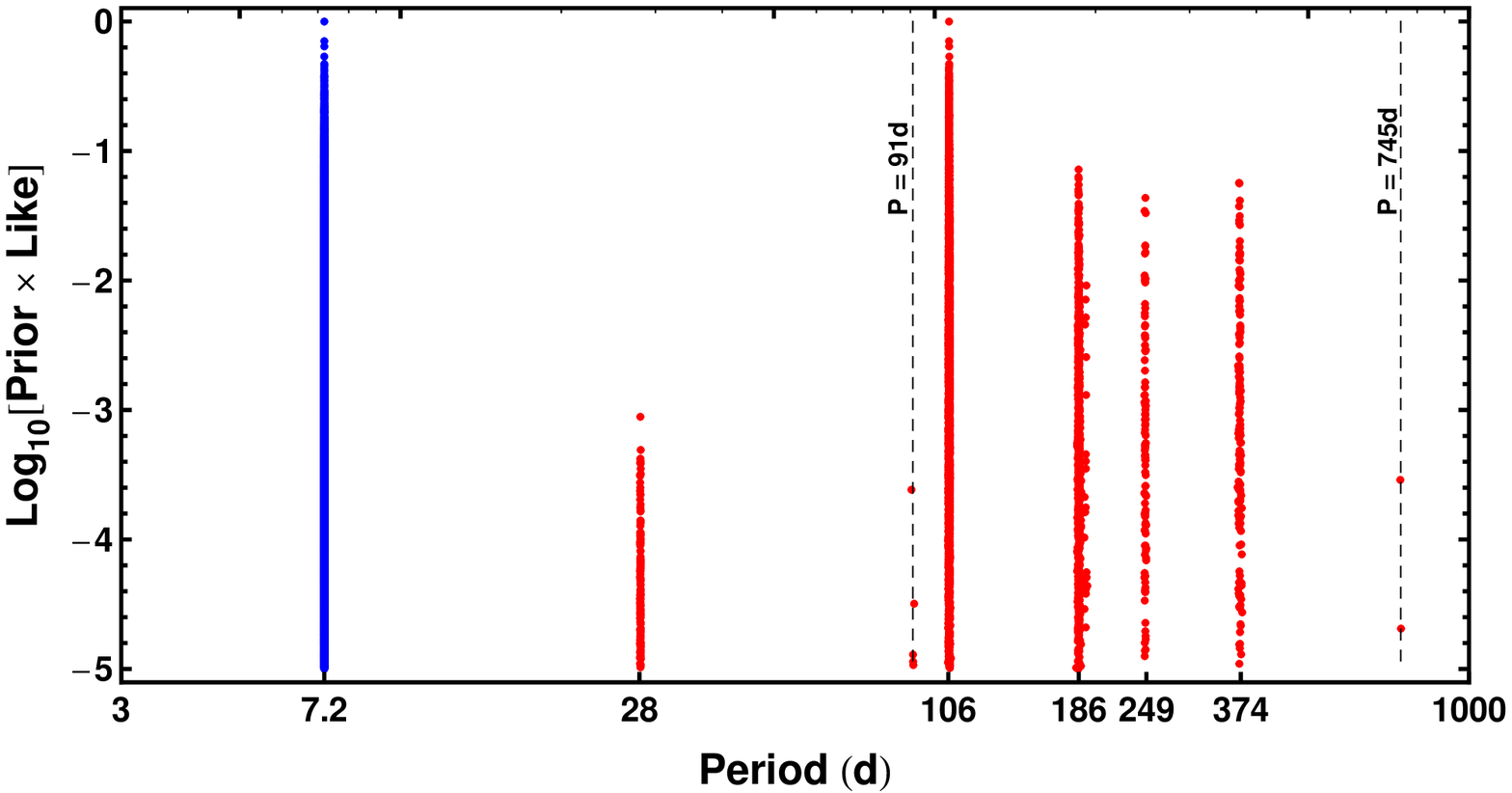}
\caption{A plot of the 2 period FMCMC parameter samples versus a normalized value of Log$_{10}$[Prior $\times$ Likelihood] for the 2 planet Kepler periodogram.}
\label{fig:2planP}
\end{figure}

\begin{figure}
\includegraphics[width=85mm]{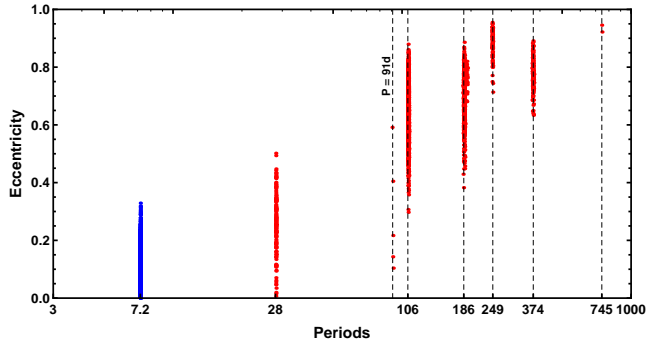}
\caption{A plot of eccentricity versus period for the FMCMC parameter samples from the 2 planet Kepler periodogram.}
\label{fig:2planEccP}
\end{figure}

Five 2 planet Kepler periodograms were computed.  In all cases signals were detected at 7.2 and 28 d but the 28 d signal was never the dominant second peak. Other periods which occured in individual periodograms included 106, 184, 249, 383 d.  Fig.~\ref{fig:2planPiter} shows the results for the periodogram which achieved the highest peak  Log$_{10}$[Prior $\times$ Likelihood]. The upper plot shows the Log$_{10}$[Prior $\times$ Likelihood] versus  iteration. The lower plot shows the two period parameters versus iteration which shows a stable 7.2 d signal while the second period transitions over many peaks thanks to the parallel tempering feature of the FMCMC algorithm. The relative peak probabilities of the second period options is illustrated in Fig.~\ref{fig:2planP} which shows a plot of the two period parameter values versus a normalized value of Log$_{10}$[Prior $\times$ Likelihood]. Fig.~\ref{fig:2planEccP} shows a plot of the eccentricity parameter values versus period. Only the 7.2 and 28.1 d periods are consistent with low eccentricity orbits. The median value of the extra noise parameter 
$s= 1.8$ m s$^{-1}$.


\subsection{Three planet model}
\label{sec:3HARPS}

\begin{figure}
\includegraphics[width=85mm]{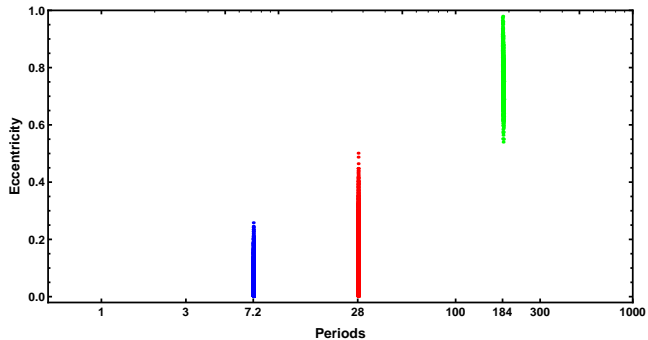}
\caption{A plot of eccentricity versus period for the FMCMC parameter samples from the 3 planet Kepler periodogram.}
\label{fig:3planEccP}
\end{figure}

Six 3 planet Kepler periodograms were computed for the data.  In all cases signals were detected at 7.2 and 28.1 d. Other periods which occured in individual periodograms were 38.8, 106, 184, 368 d. In three cases the third period was 184 d. Fig.~\ref{fig:3planEccP} shows a plot of eccentricity versus period for the periodogram which achieved the highest peak  Log$_{10}$[Prior $\times$ Likelihood]. The 7.2 and 28.1 d signals are consistent with low eccentricity orbits. The third period invariably had a high eccentricity as is the case for the 184 d period in the figure. The median value of the extra noise parameter $s= 1.4$ m s$^{-1}$.

\subsection{Four planet model}
\label{sec:4HARPS}

\begin{figure}
\includegraphics[width=85mm]{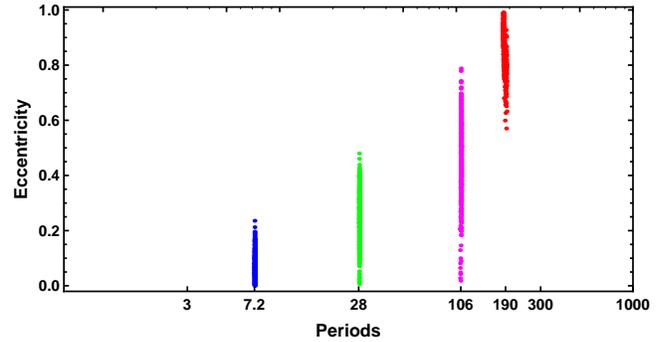}
\caption{A plot of eccentricity versus period for the FMCMC parameter samples from the 4 planet Kepler periodogram..}
\label{fig:4planEccP}
\end{figure}

Three 4 planet Kepler periodograms were computed for the data. In all cases signals were detected at 7.2, 28.1, 106, \& 190 days. Additional periods that were detected in different runs were  38.8 and 91d. Fig.~\ref{fig:4planEccP} shows a plot of eccentricity versus period for the periodogram which achieved the highest peak  Log$_{10}$[Prior $\times$ Likelihood]. The median value of the extra noise parameter $s= 1.0$ m s$^{-1}$.

\subsection{Five planet model}
\label{sec:5HARPS}

\begin{figure}
\includegraphics[width=85mm]{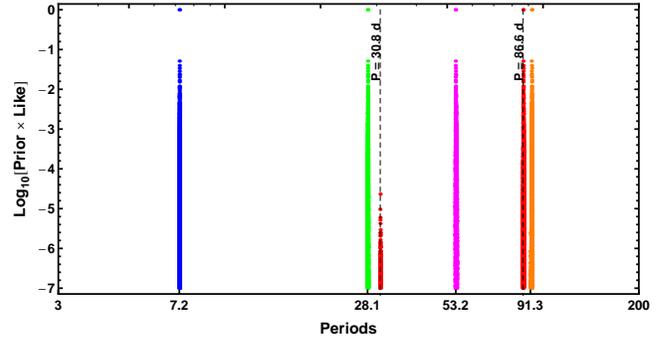}
\caption{A plot of the five period parameter values versus a normalized value of Log$_{10}$[Prior $\times$ Likelihood] for the 5 planet Kepler periodogram.}
\label{fig:5planP}
\end{figure}

\begin{figure}
\includegraphics[width=85mm]{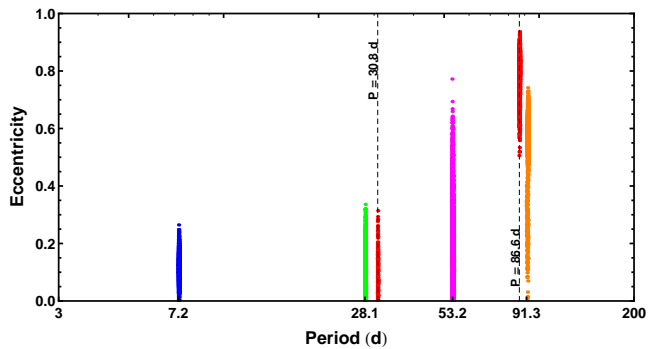}
\caption{A plot of eccentricity versus period for the FMCMC parameter samples from the 5 planet Kepler periodogram..}
\label{fig:5planEccP}
\end{figure}

Four 5 planet Kepler periodograms were computed for the data. In all cases signals were detected at 7.2, 28.1, 53.2, \& 91 d. Other periods which occured in individual periodograms included  30.8, 38.8, 87, 106, 128, 184 d. In some cases the fifth period exhibited a variety of periods. Fig.~\ref{fig:5planP} shows the periodogram which achieved the highest peak  Log$_{10}$[Prior $\times$ Likelihood]. In this case the extra period exhibits a dominant peak at 86.6 d and a weaker peak at 30.8 d. Fig.~\ref{fig:5planEccP} is a plot of eccentricity versus period. The 86.6 d signal exhibits a high eccentricity while the 30.8 d signal has a low eccentricity. The median value of the extra noise parameter $s= 0.79$ m s$^{-1}$.

\subsection{Six planet model}
\label{sec:6HARPS}

\begin{figure}
\includegraphics[width=85mm]{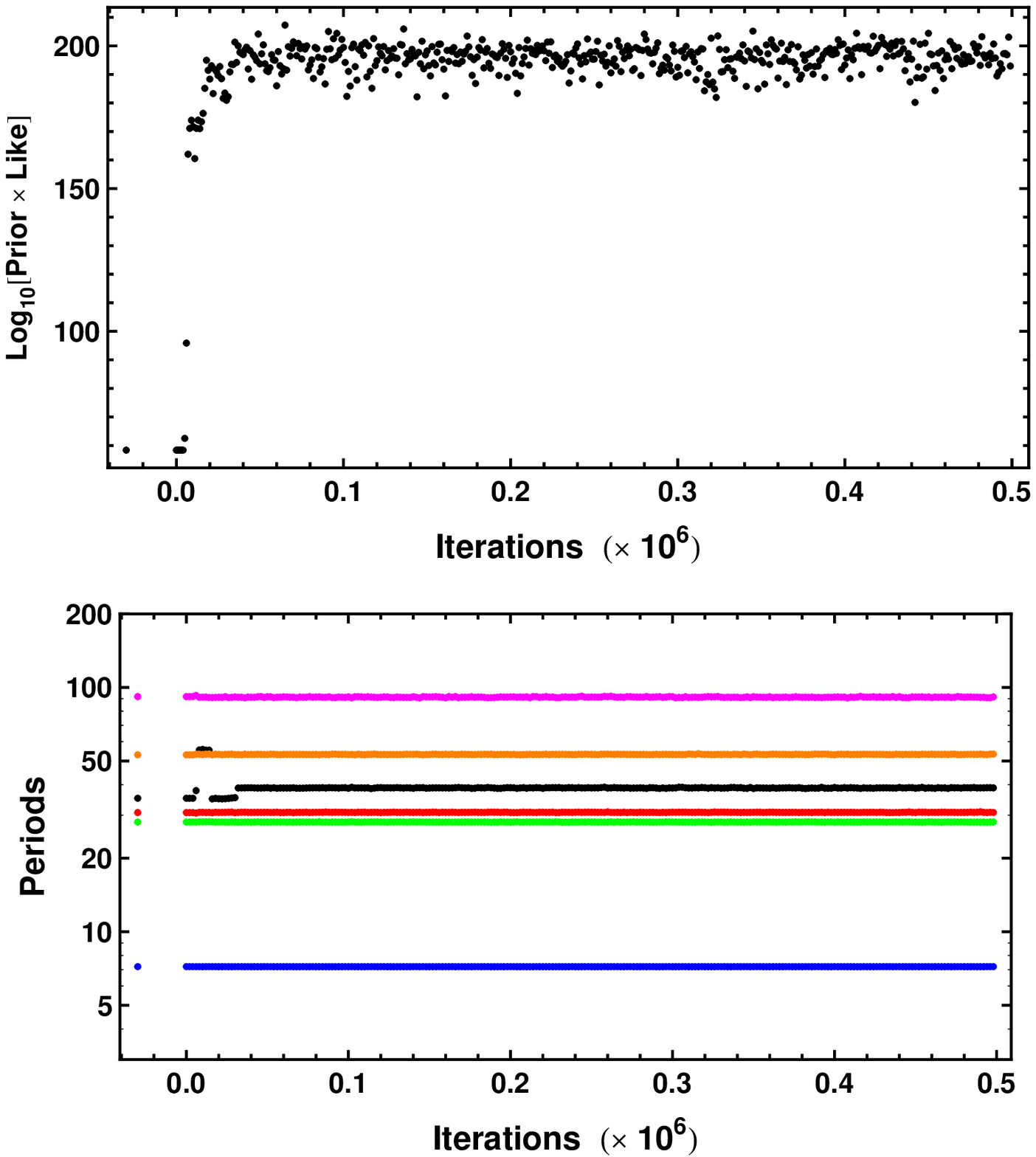}
\caption{The upper panel is a plot of the Log$_{10}$[Prior $\times$ Likelihood] versus iteration for the six planet fit of the HARPS data. The lower shows the values of the six unknown period parameters versus iteration number. The six starting periods are shown on the left hand side of the plot at a negative iteration number.}
\label{fig:6planPiter}
\end{figure}

We next computed a 6 planet Kepler periodogram analysis of the data. An initial blind search with starting periods of 5, 10, 15, 20, 50, 100 d yielded 5 well defined periods at 7.2, 28.1, 30.8, 53.2, 91.3 d. The remaining period was split between a high eccentricity 14.4 d period and a low eccentricity 35 d period.  A second run starting from the 5 well defined periods plus the 35 d low eccentricity option  is shown in Fig.~\ref{fig:6planPiter}. After a temporary stay at the 35 d period, the black trace transitions to a stable 38. 8 d peak~\footnote{Another 6 planet Kepler periodogram was computed starting from the best 6 planet parameters set with the exception that 53.2 d period which was replaced by twice this period which corresponds roughly with the assumed rotation period of the star of 105 d (\citealt{Delfosse2012}, based on a periodogram of a stellar activity diagnostic). This solution with periods of 7.2, 28.1, 30.8, 38.8, 91 and 106.5 d had a peak probability 2600 times lower.}. As discussed in Section~\ref{sec:modsel} on model selection, the Bayes factor favors the 6 planet model by a factor of 137 compared to the next leading contender. The median value of the extra noise parameter $s= 0.54$ m s$^{-1}$.

\begin{figure}
\includegraphics[width=85mm]{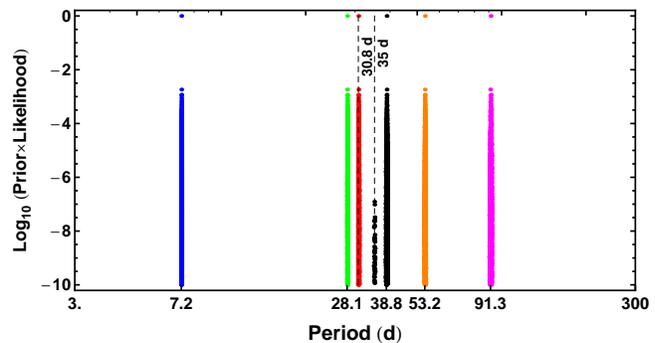}
\caption{Full HARPS data set: A plot of the 6 period parameter values versus a normalized value of Log$_{10}$[Prior $\times$ Likelihood] for the 6 planet Kepler periodogram (includes burn-in points to show the 35 d signal). The fourth period, shown in black, exhibits two peaks one at 35 d and the other at 38.8 d. The 35 d period is a one year alias of the stronger 38.8 d period.}
\label{fig:6planP}
\end{figure}

\begin{figure}
\includegraphics[width=85mm]{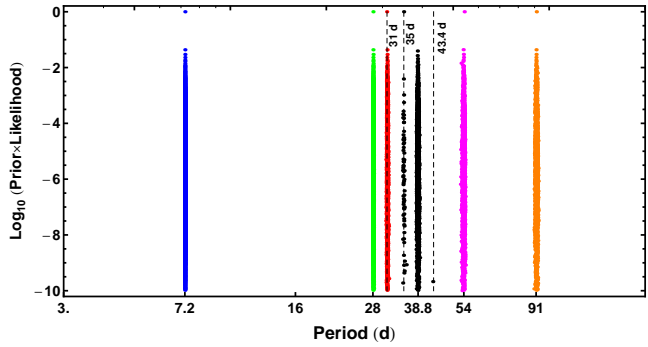}
\caption{Subset of HARPS data (first 143 points): A plot of the 6 period parameter values versus a normalized value of Log$_{10}$[Prior $\times$ Likelihood] for the 6 planet Kepler periodogram (includes burn-in points to show the 35 d signal). The fourth period, shown in black, exhibits two peaks one at 35 d and the other at 38.8 d. For the partial data set the 35 d alias has a higher peak probability density. The other one year alias at 43.4 d is just discernible.}
\label{fig:6planP143}
\end{figure}

Fig.~\ref{fig:6planPiter} shows a plot of the 6 period parameter values (including burn-in points to show the 35 d signal) versus a normalized value of Log$_{10}$[Prior $\times$ Likelihood] for the 6 planet  Kepler periodogram. The fourth period, shown in black, exhibits two peaks, a weak one at 35 d and the other at 38.8 d. The 35 d period coincides with a one year alias of the stronger 38.8 d period. The spectral window function for the HARPS data exhibits two peaks, 1 d and one year.  We can gain further insight into the relationship of the 35 and 38.8 d periods from a 6 planet Kepler periodogram of a subset of the HARPS data, i.e., the first 143 data points. Again during the burn-in, the fourth period locks on to the 35 d peak before transitioning to the 38.8 d peak. Fig.~\ref{fig:6planP143} shows the 6 period parameter values (including burn-in points) versus a normalized value of Log$_{10}$[Prior $\times$ Likelihood] for the 143 d sub-sample. In this case, the alias at 35 d actually has a higher peak probability density by a factor of $\sim 10$, even though the much larger number of samples for the 38.8 d peak indicates that there is more probable associated with the 38.8 d peak. A narrow peak in the joint parameter space of high probability density can contain much less total probability than a broader region of lower probability density.  Note that the other one year alias of the 38.8 d period at 43.4 d coincides with the location of a very weak feature. Comparing Fig's~\ref{fig:6planP} and ~\ref{fig:6planP143}, it is clear that the addition of the extra data points in the latest HARPS sample of \cite{Delfosse2012} has suppressed the alias at 35 d by a factor of $\sim 10^8$.
\begin{figure}
\includegraphics[width=85mm]{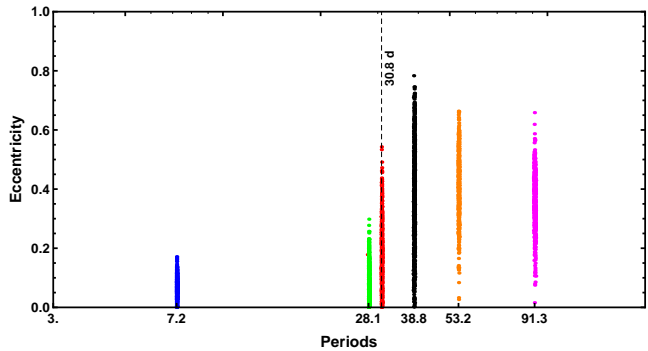}
\caption{Full HARPS data set: A plot of eccentricity versus period for the 6 planet Kepler periodogram for the post-burn-in FMCMC samples.}
\label{fig:6planEccP}
\end{figure}

It was not until the 6 planet model that the 30.8  and 38.8 d signals became the dominant third and fourth periods. Both are consistent with low eccentricity orbits as shown in Fig.~\ref{fig:6planEccP} which is a plot of eccentricity versus period for the post-burn-in samples for the full HARPS data set. 

\begin{figure}
\includegraphics[width=85mm]{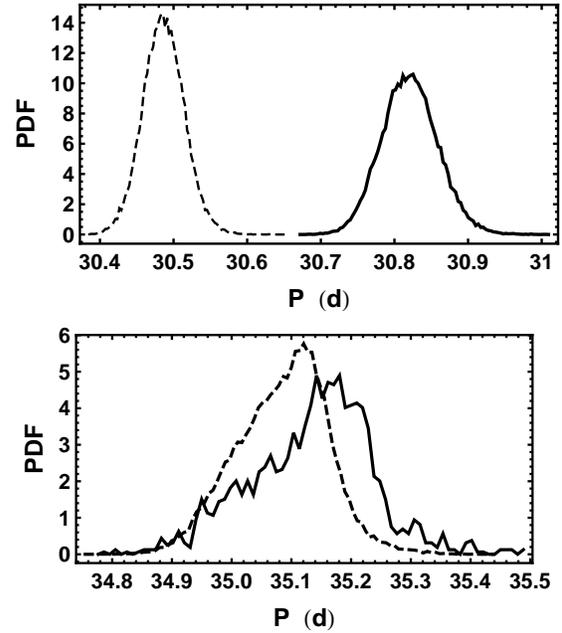}
\caption{The upper panel shows a comparison of the predicted marginal distribution (dashed) of the closest one year alias of the 28.14 d signal to the marginal distribution for the 30.82 d signal (solid). For comparison, the bottom panel shows the predicted marginal distribution of the lower one year alias (dashed) of the 38.8 d signal to the marginal distribution obtained for the 35 d signal (solid) which was observed during the burn-in phase of the full HARPS data set.}
\label{fig:6planPalias}
\end{figure}

The closeness of the 28.1 and 30.8 day periods raises questions about a possible relationship between these two signals. Could the 30.8 d signal be an alias of the 28.1 d signal? An alias results from a convolution of the spectral window function of the data sample times with the spectrum of the real signal. If the real signal is removed then an alias of that signal should not be present in the residuals. In our case the multi-planet model is requiring both to be present. Nevertheless, we transformed the marginal distribution of the second period, $P_2$ (28.14 d signal), by the relationship $P_{\rm alias} = 1/(1/P_2 - 1/365.25)$ to obtain the pridicted marginal distribution the one year alias. This is compared it to the marginal for $P_3$ (30.82 d signal) in the top panel of Fig.~\ref{fig:6planPalias}. Clearly, the predicted alias (shown dashed) does not overlap the marginal for $P_3$. For comparison, the bottom panel shows the predicted marginal distribution of the lower one year alias (dashed) of the 38.8 d signal to the marginal distribution obtained for the 35 d signal (solid) which was observed during the burn-in phase of the full HARPS data set. In this case the distributions coincide.  

\begin{figure}
\includegraphics[width=85mm]{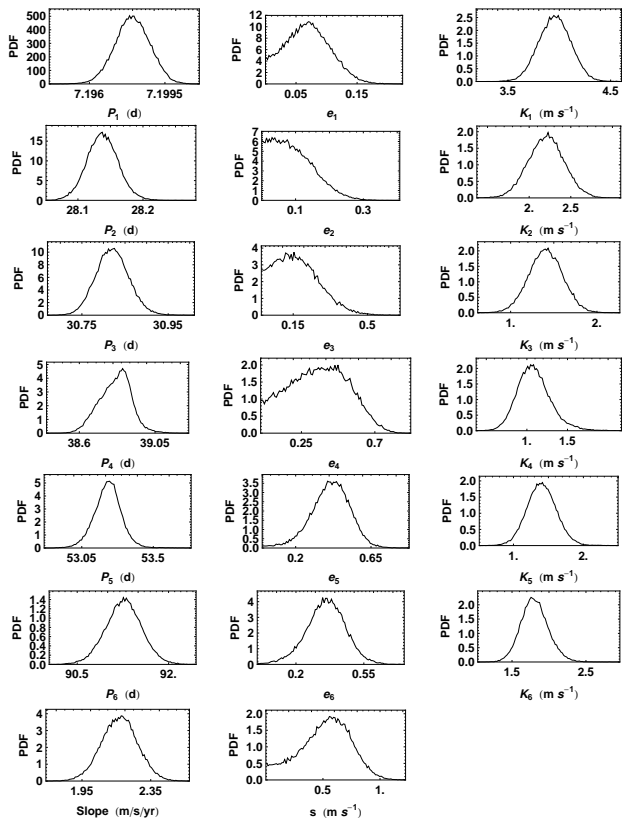}
\caption{A plot of a subset of the FMCMC parameter marginal distributions for the 6 planet Kepler periodogram.}
\label{fig:6planMarg}
\end{figure}

Fig.~\ref{fig:6planMarg} shows a plot of a subset of the FMCMC parameter marginal distributions for the 6 planet FMCMC fit of the data. The eccentricities of the 4 lowest periods are consistent with low eccentricity orbits while the 53.2 and 91.3 d periods show peak eccentricities of 0.41 and 0.36, respectively. The median value of extra noise parameter $s= 0.54$ m s$^{-1}$.

\begin{figure}
\includegraphics[width=85mm]{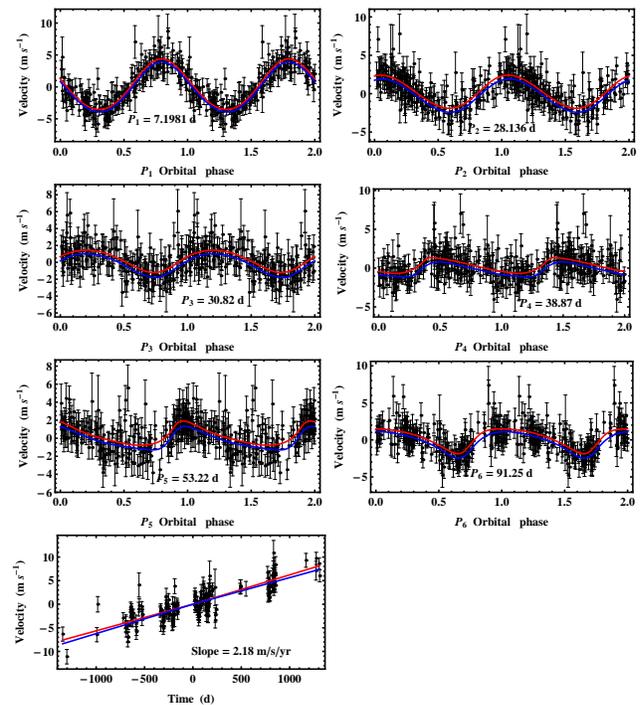}
\caption{The top left panel shows the data and model fit versus 7.2 d period phase after removing the effects of the 5 other orbital periods plus $V$ and $slope$ parameters. The upper and lower curves are the mean FMCMC model fit $\pm 1$ standard deviation. The next five panels correspond to phase plot for the other five periods. The bottom panel is for the $slope$ parameter after removing the effects of the 6 orbital periods plus $V$. }
\label{fig:6planPhasePls}
\end{figure}

Phase plots for the 6 planet model are shown in Fig.~\ref{fig:6planPhasePls}. The top left panel shows the data and model fit versus 7.2 d period phase after removing the effects of the five other orbital periods plus V and $slope$ parameter. 
The FMCMC output for each iteration is a vector of the 6 planet orbital parameter set plus $V$, $slope$, and the extra noise parameters. The 7.2 d period phase plot is constructed from a sample of FMCMC iterations (typically 300) and for each iteration we compute the predicted velocity points for that realization of the 5 planet plus $V$ and $slope$ parameters. We then construct the mean of these model predictions and subtract the mean prediction from the data points. These residuals for the set of observation times are converted to residuals versus phase using the mode of the marginal distribution for the 7.2 d period parameter. A period phase model velocity fit is then computed at 100 phase points for each realization of the 7.2 d planet parameter set obtained from the same sample of 300 FMCMC iterations. At each of these 100 phase points we construct the mean model velocity fit and mean $\pm 1$ standard deviation. The upper and lower solid curves in Fig.~\ref{fig:6planPhasePls} are the mean FMCMC model fit $\pm 1$ standard deviation. Thus, 68.3\% of the FMCMC model fits fall between these two curves. 

The next five panels correspond to phase plot for the other 5 periods. In each panel the quoted period is the mode of the marginal distribution.  The bottom panel is for the $slope$ parameter after removing the effects of the 6 orbital periods plus $V$.

\begin{figure}
\includegraphics[width=85mm]{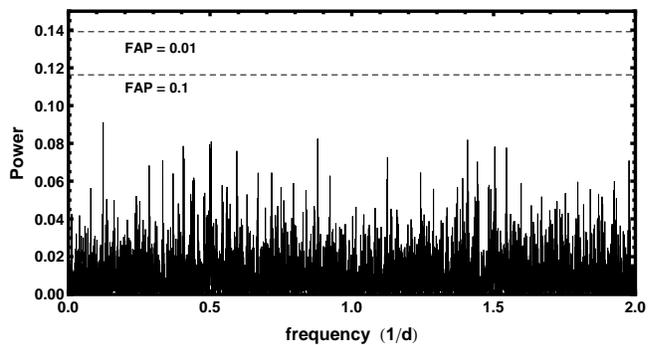}
\caption{A periodogram of the 6 planet fit residuals.}
\label{fig:6planResidPeriodogram}
\end{figure}

Fig.~\ref{fig:6planResidPeriodogram} shows a generalized Lomb-Scargle (GLS) periodogram \citep{Zechmeister2009} for the maximum {\it a posteriori} (MAP) parameter values of the 6 planet fit residuals. The GLS allows for a floating offset and weights. The dashed horizontal lines correspond to peak periodogram levels for which the false alarm probability (FAP) would $= 0.1\ \& \ 0.01$~\footnote{The interesting region of power is where the frequentist p-value is small ($\ll 1$). In this region the FAP is given approximately by FAP$  \approx M * (1-z)^{(N-3)/2}$ \citep{Zechmeister2009}, where $z =$ maximum periodogram power, $M =$ the number of independent frequencies and $N$ is the number of data points.  Clearly, the FAP value for the actual highest peak in the periodogram is much larger than 0.1. \cite{Cumming2004} recommends setting $M = \Delta f/\delta f$, where $\Delta f$ is the frequency range examined $\approx f_{\rm max}$, and $\delta f =$ the resolution of the periodogram $\approx 1/T$ where $T$ is the duration of the data.}. There is no evidence of any significant peaks or red noise in these residuals. 

\begin{figure}
\begin{center}
\includegraphics[width=85mm]{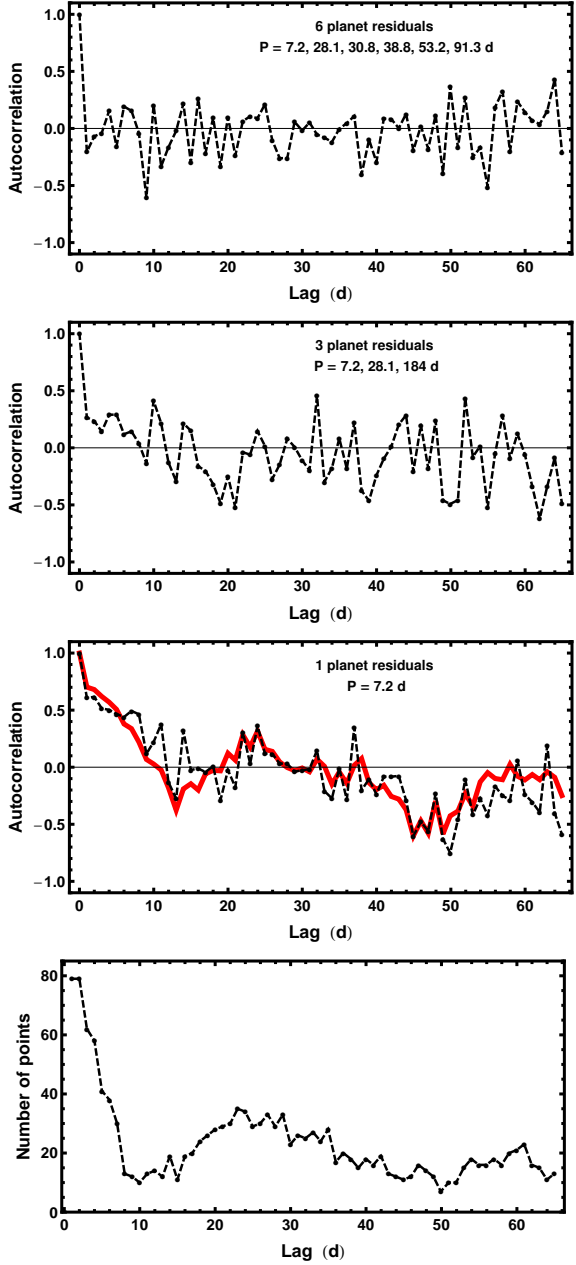}
\end{center}
\caption{The top panel shows the autocorrelation function, $\rho(j)$, of the HARPS 6 planet fit residuals computed as explained in the text. The next two panels show $\rho(j)$ for the 3 planet residuals and the 1 planet residuals (together with a simulation thick solid line). The bottom panel shows the number of HARPS sample pairs available for computing the autocorrelation function versus lag.}
\label{fig:ACF}
\end{figure}

Another way of exploring the residuals is to compute the autocorrelation function, $\rho(j)$. Fig.~\ref{fig:ACF} shows $\rho(j)$ of the residuals for the 6, 3 and 1 planet fits computed from equation~(\ref{eq:ACF1}).
\begin{equation}
\rho(j) = \frac{\sum_{\rm overlap} [(x_{i}-\overline{x})\ (x_{i+j}-\overline{x})]}{\sqrt{\sum_{\rm overlap} (x_{i}-\overline{x})^2} \times \sqrt{\sum_{\rm overlap} (x_{i+j}-\overline{x})^2}},
\label{eq:ACF1}
\end{equation}
where $x_i$ is the $i^{\rm th}$ residual, $j$ is the lag and $\overline{x}$ is the mean of the samples in the overlap region. Because the data are not uniformly sampled, for each lag all sample pairs that differed in time by this lag $\pm 0.1$ d were utilized. The bottom right panel of Fig.~\ref{fig:ACF} shows a plot of the number of such sample pairs as a function of the lag. Clearly for large lags the uncertainty in computed $\rho(j)$ is expected to be larger. 

Clearly, the 6 planet model autocorrelation function is consistent with white noise. The thick solid line in the 1 planet residuals panel is the average autocorrelation function generated from 400 simulated data sets of a 5 planet model (28.1, 30.8, 38.8, 53.2, and 91.4 d periods) together with the measurement errors. The 5 planet model parameter values are the MAP values derived from the 6 planet Kepler periodogram of the real data. The good agreement between the simulation and actual $\rho(j)$ argues that in the case of Gl 667C any colored noise evident in the 1 planet fit is nicely accounted for by additional signals (mostly planets) not included in the model.

\subsection{Seven planet model}
\label{sec:7HARPS}

\begin{figure}
\includegraphics[width=85mm]{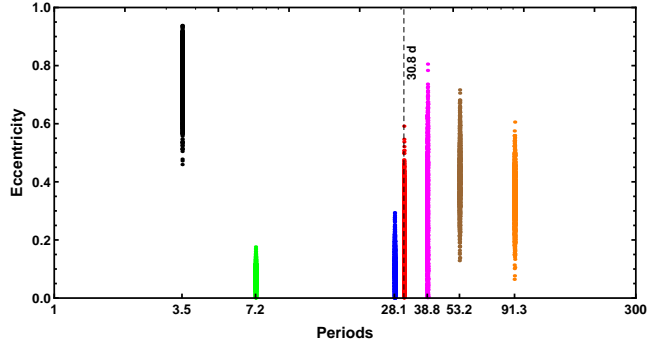}
\caption{A plot of eccentricity versus period for the 7 planet Kepler periodogram of the full HARPS data.}
\label{fig:7planEccP}
\end{figure}

In spite of the absence of any obvious signal  from the 6 planet residual periodogram, we computed three 7 planet Kepler periodogram to see if there were any new surprises. Each trial used a starting period set of 7.2, 28.1, 30.8, 35.2, 53.2, 91.3 d  and a different choice for the seventh period. Encouragingly, all trials recovered the 6 periods found in the 6 planet fit which indicates they are stable features. Fig.~\ref{fig:7planEccP} shows the plot of eccentricity versus period for the trial which achieved the highest peak  Log$_{10}$[Prior $\times$ Likelihood]. The additional 3.5 d period is a weak high eccentricity feature typical of noise \citep{Gregory2010}. The median value of the extra noise parameter $s= 0.23$ m s$^{-1}$. 

\section{Model Selection}
\label{sec:modsel}

One of the great strengths of Bayesian analysis is the built-in Occam's razor. More complicated models contain larger numbers of parameters and thus incur a larger Occam penalty, which is automatically incorporated in a Bayesian model selection analysis in a quantitative fashion (see for example, \cite{Gregorybook}, p. 45). The analysis yields the relative probability of each of the models explored.

To compare the posterior probability of the i$^{\rm th}$ planet model to the 4 planet model we need to evaluate the odds ratio,
$O_{i4} =p(M_{i} | D,I)/p(M_{4} | D,I)$, the ratio of the posterior probability
of model $M_{i}$ to model $M_{4}$.  Application of Bayes 
theorem leads to,
\begin{equation}
O_{i4} = {p(M_{i} | I) \over p(M_{4} | I)}\;
      {p(D | M_{i},I) \over p(D | M_{4},I)}
       \equiv {p(M_{i} | I) \over p(M_{4} | I)}\; B_{i4}
\label{eq:orbit22}
\end{equation}
where the first factor is the prior odds ratio, and the second factor
is called the {\it Bayes factor}, $B_{i4}$. The Bayes factor is the ratio of
the marginal (global) likelihoods of the models. The marginal likelihood for model $M_i$ is given by
\begin{equation}
p(D|M_i,I)= \int d\vec{X} p(\vec{X}|M_i,I)\times p(D|\vec{X},M_i,I).
\label{eq:marglike}
\end{equation}
Thus Bayesian model selection relies on the ratio of marginal likelihoods, not maximum likelihoods. The marginal likelihood is the weighted average of the conditional likelihood, weighted by the prior probability distribution of the model parameters.

The marginal likelihood can be expressed as the product of the maximum likelihood and the Occam penalty (see \cite{Gregorybook}, page 48). The Bayes factor will favor the more complicated model only if the maximum likelihood ratio is large enough to overcome this penalty. In the simple case of a single parameter with a uniform prior of width $\Delta X$, and a centrally peaked likelihood function with characteristic width $\delta X$, the Occam factor is $\approx \delta X/\Delta X$. If the data is useful then generally $\delta X \ll \Delta X$. For a 
model with $m$ parameters, each parameter will contribute a term to the overall Occam penalty. The Occam penalty depends not only on the number of parameters but also on the prior range~\footnote{The more surprising the result the stronger the evidence required to overcome our skepticism.} of each parameter (prior to the current data set, $D$), as symbolized in this simplified discussion by $\Delta X$. If two models have some parameters in common then the prior ranges for these parameters will cancel in the calculation of the Bayes factor. 

To make good use of Bayesian model selection, we need to fully specify priors that are independent of the current data $D$. In most instances we are not particularly interested in the Occam factor itself, but only in the relative probabilities of the competing models as expressed by the Bayes factors. Because the Occam factor arises automatically in the marginalization procedure, its effect will be present in any model selection calculation. Note: no Occam factors arise in parameter estimation problems. Parameter estimation can be viewed as model selection where the competing models have the same complexity so the Occam penalties are identical and cancel out. 
   
The MCMC algorithm produces samples which are in proportion to the posterior probability distribution which is fine for parameter
estimation but one needs the proportionality constant for estimating the model marginal likelihood. 
\citet{Clyde2007} reviewed the state of techniques for model selection from a statistical perspective and \citet{FordGregory2007} have evaluated the performance of a variety of marginal likelihood estimators in the exoplanet context. 

\begin{table*}
  \caption{Marginal likelihood estimates, Bayes factors relative to model 4, and false alarm probabilities. The last two columns list the median of extra noise parameter, $s$, and the RMS residual. The MAP value of $s$ appears below in parentheses.}
  \label{tab:modelSel}
  \begin{tabular}{@{}llllllll@{}}
  \hline
   Model & Periods &  Marginal & Bayes factor & False Alarm & $s$ &RMS residual \\
         & (d) &  Likelihood & \ \ nominal & Probability &  (m s$^{-1})$ & (m s$^{-1}$)\\
\hline
$M_{0}$ & & $ 5.69\times 10^{-223}$ & $1.1\times 10^{-35}$ & & 3.6 & 3.9\\
& & & & &(3.6) &\\
& & & & & &\\
$M_{1}$ & $(7.2)$& $(8.57_{-0.04}^{+0.07})\times 10^{-202}$ & $1.6\times 10^{-14}$ & $6.6\times 10^{-22}$ & 2.3 & 2.6\\
& & & & &(2.2) &\\
& & & & & &\\
$M_{2}$ & $(7.2,106)$& $(2.37_{-0.09}^{+0.06})\times 10^{-196}$ & $6.5\times 10^{-9}$ & $3.6\times 10^{-6}$ & 1.8 & 2.3\\
& & & & &(1.6) &\\
& & & & & &\\
$M_{3}$ & $(7.2,28.1,184)$ & $(1.19_{-0.09}^{+0.25}) \times 10^{-190}$ & $2.28\times 10^{-2}$ & $2.0\times 10^{-6}$ & 1.4 & 2.0\\
& & & & &(1.3) &\\
& & & & & &\\
$M_{4}$  &$(7.2,28.1,106,190)$& $(5.22_{\times 0.38}^{\times 2.5}) \times 10^{-188}$ & $1.0$ & $2.3\times 10^{-3}$ & 1.0 & 1.8\\
& & & & &(0.71) &\\
& & & & & &\\
$M_{5}$  &$(7.2,28.1,53,86.6,91)$& $(3.33_{\times 0.63}^{\times 1.7}) \times 10^{-188}$ & $0.64$ & $0.61$ & 0.79 & 1.6\\
& & & & &(0.70) &\\
& & & & & &\\
$M_{6}$  &$(7.2,28.1,30.8,38.8,53,91)$& $(7.2_{\times 0.71}^{\times 2.0}) \times 10^{-186}$ & $137$ & $0.012$ & 0.54 & 1.5\\
& & & & &(0.08) &\\
& & & & & &\\
$M_{7}$  &$(3.5,7.2,28.1,30.8,38.8,53,91)$& $(5.02_{\times 0.36}^{\times 4.2}) \times 10^{-188}$ & $0.96$ & $0.993$ & 0.23 & 1.4\\
& & & & & (0.03) &\\
\hline
\end{tabular}
\end{table*}
Estimating the marginal likelihood is a very big challenge for models with large numbers of parameters, e.g., our seven planet model has 38 parameters. In this work we employ the nested restricted Monte Carlo (NRMC) method introduced in \cite{Gregory2010} and described in more detail  in \cite{Gregory2012} to estimate the marginal likelihoods. Monte Carlo (MC) integration can be very inefficient in exploring the whole prior parameter range because it randomly samples the whole volume. The fraction of the prior volume of parameter space containing significant probability rapidly declines as the number of dimensions increase. For example, if the fractional volume with significant probability is 0.1 in one dimension then in 38 dimensions the fraction might be of order $10^{-38}$. In restricted MC integration (RMC) this is much less of a problem because the volume of parameter space sampled is greatly restricted to a region delineated by the outer borders of the marginal distributions of the parameters for the particular model. 

In RMC, most of the random samples occur close to the outer borders of the restricted region because the contribution to the volume of parameter space is greatest there. In NRMC integration, multiple boundaries are constructed based on credible regions ranging from 30\% to $\ge 99\%$, as needed. We are then able to compute the contribution to the total integral from each nested interval and sum these contributions. For example, for the interval between the 30\% and 60\% credible regions, we generate random parameter samples within the 60\% region and reject any sample that falls within the 30\% region. Using the remaining samples we can compute the contribution to the NRMC integral from that interval. 

\begin{figure*}
\includegraphics[width=160mm]{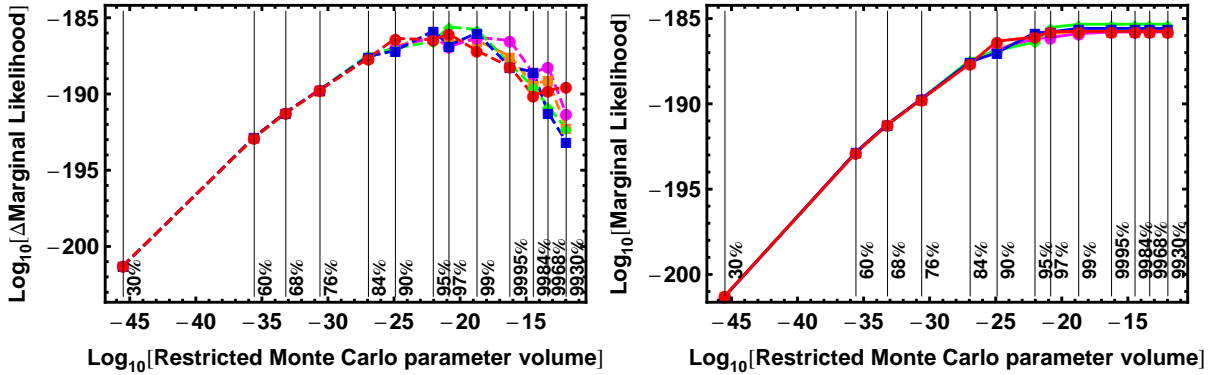}
\caption{Left panel shows the contribution of the individual nested intervals to the NRMC marginal likelihood for the 6 planet model. The right panel shows the integral of these contributions versus the parameter volume of the credible region.}
\label{fig:6planNestedRMC}
\end{figure*}

The left panel of Fig.~\ref{fig:6planNestedRMC} shows the contributions from the individual intervals for 5 repeats of the NRMC evaluation for the 6 planet model. Note the large range in parameter volume on the abscissa. The right panel shows the summation of the individual contributions versus the volume of the credible region. The credible region listed as 9995\% is defined as follows. Let $X_{U99}$ and $X_{L99}$ correspond to the upper and lower  boundaries of the 99\% credible region, respectively, for any of the parameters. Similarly, $X_{U95}$ and $X_{L95}$ are the upper and lower boundaries of the 95\% credible region for the parameter. Then $X_{U9995} = X_{U99}+(X_{U99}-X_{U95})$ and $X_{L9995} = X_{L99}+(X_{L99}-X_{L95})$. Similarly, $X_{U9984} = X_{U99}+(X_{U99}-X_{U84})$. For each credible region interval approximately
320,000 MC samples were used. 

The mean value of the prior $\times$ likelihood within the 30\% credible region is a factor of $4.8 \times 10^{12}$ larger than the mean in the shell between
the 97 and 99\% credible regions. However, the volume of parameter space in the shell between the 97 and 99\% credible regions is a factor of $ 5.9 \times 10^{26}$ larger than
the volume within the 30\% credible region so the contribution from the latter to the
marginal likelihood is negligible. For further details on the NRMC method see \cite{Gregory2012}.

The NRMC method is expected to underestimate the marginal likelihood in high dimensions and this underestimate is expected to become worse the larger the number of model parameters, i.e. increasing number of planets (\citealt{Gregory2012}). When we conclude, as we do, that the NRMC computed odds in favor of the six planet model compared to the four planet model is $137$ we mean that the true odds is $\ge 137$. Thus the NRMC method is conservative. One indication of the break down of the NRMC method is the increased spread in the results for repeated evaluations. 
 
We can readily convert the Bayes factors to a Bayesian False Alarm Probability (FAP) which we define in equation~\ref{eq:FAP1}. For example, in the context of claiming the detection of $m$ planets the FAP$_m$ is the probability that there are actually fewer than $m$ planets, i.e., $m-1$ or less.
\begin{equation}
{\rm FAP}_m = \sum_{i=0}^{m-1}({\rm prob. of \ i \ planets}) 
\label{eq:FAP1}
\end{equation}

If we assume {\it a priori} that all models under consideration are equally likely, then the probability of each model is related to the Bayes factors by
\begin{equation}
p(M_i\mid D,I) = {{B_{i4}} \over {\sum_{j=0}^{N} B_{j4}}}
\label{eq:FAP2}
\end{equation}
where $N$ is the maximum number of planets in the hypothesis space under consideration, and of course
$B_{44} = 1$. For the purpose of computing FAP$_m$ we set $N = m$. Substituting Bayes factors from Table~\ref{tab:modelSel} into equation~\ref{eq:FAP1} gives
\begin{eqnarray}
{\rm FAP}_6 & =& {{(B_{04} + B_{14} + B_{24}+ B_{34}+ B_{44}+ B_{54})}\over {\sum_{j=0}^{6} B_{j4}}}=0.012\nonumber\\
\label{eq:FAP3}
\end{eqnarray}
For the 6 planet model we obtain a low FAP $\approx 10^{-2}$. 

Table~\ref{tab:modelSel} gives the NMRC Marginal likelihood estimates, Bayes factors and false alarm probabilities for 0, 1, 2, 3, 4, 5, 6, and 7 planet models which are designated $M_0, \cdots, M_7$. The last two columns list the median estimate of the extra noise parameter, $s$ (MAP value in parentheses below), and the RMS residual. For each model the NRMC calculation was repeated 5 times and the quoted errors give the spread in the results, not the standard deviation. The Bayes factors that appear in the third column are all calculated relative to model 4. 

A summary of the 6 planet model parameters and their uncertainties are given in Table~\ref{tab:parerrorsM6}. The quoted value is the median of the marginal probability distribution for the parameter in question and the error bars identify the boundaries of the 68.3\% credible region~\footnote{In practice, the probability density for any parameter $X$  is represented by a finite list of values $p_i$ representing the probability in discrete intervals $\delta X$. A simple way to compute the 68.3\% credible region, in the case of a marginal with a single peak, is to sort the $p_i$ values in descending order and then sum the values until they approximate 68.3\%, keeping track of the upper and lower boundaries of this region as the summation proceeds.}. The value immediately below in parenthesis is the MAP estimate, the value at the maximum of the joint posterior probability distribution.  For the eccentricity parameter we also include the mode within double parentheses. It is not uncommon for the MAP estimate to fall close to the borders of the credible region. 
In one case, the eccentricity of the fourth planet, the MAP estimate falls just outside the 68.3\% credible region which is one reason why we prefer to quote median or mode values as well. 
The semi-major axis and $M \sin i$ values are derived from the model parameters assuming a stellar mass of $0.31\pm0.019$ M$_{\sun}$ \citep{Anglada2012a}. The quoted errors on the semi-major axis and $M \sin i$ include the uncertainty in the stellar mass. 

\begin{table*}
 \centering
 \begin{minipage}{140mm}
  \caption{Six planet model parameter estimates. The value immediately below in parenthesis is the MAP estimate. For the eccentricity parameter, the value within double parentheses is the mode.}
  \label{tab:parerrorsM6}
  \begin{tabular}{@{}lllllll@{}}
  \hline
   Parameter  & planet 1 & planet 2 & planet 3 & planet 4 & planet 5 & planet 6 \\
\hline
$P$  (d) & $7.1980_{-.0008}^{+.0009}$ & $28.138_{-.0023}^{+.0023}$& $30.82_{-.04}^{+.04}$& $38.82_{-0.09}^{+0.09}$& $53.22_{-0.08}^{+0.08}$ & $91.26_{-0.28}^{+0.30}$  \\
& (7.1972)& (28.120)& (30.82) & (38.87) & (53.23)  & (91.37) \\
& & & & & &\\
$K$ (m s$^{-1}$) & $3.96_{-0.17}^{+0.16}$ & $2.21_{-0.22}^{+0.21}$ & $1.41_{-0.20}^{+0.19}$ & $1.07_{-0.20}^{+0.18}$ & $1.40_{-0.21}^{+0.20}$& $1.80_{-0.18}^{+0.18}$  \\
& (3.95) & (2.32) & (1.45)& (1.33)& (1.55) & (1.76) \\
& & & & & &\\
$e$ & $0.068_{-0.039}^{+0.037}$ & $0.083_{-0.081}^{+0.035}$  & $0.16_{-0.15}^{+0.08}$ & $0.35_{-0.21}^{+0.21}$ & $0.41_{-0.10}^{+0.12}$& $0.36_{-0.10}^{+0.10}$  \\
& (0.068) & (0.123) & (0.22) & (0.58) & (0.48)& (0.34) \\
& ((0.068)) & ((0.028)) & ((0.115)) & ((0.48)) & ((0.45))& ((0.35)) \\
& & & & & &\\
$\omega$  (deg) & $342_{-37}^{+35}$ & $269_{-65}^{+68}$ &  $209_{-50}^{+48}$ &  $284_{-32}^{+31}$ &  $305_{-23}^{+20}$ &  $219_{-22}^{+20}$ \\
& (338) & (266) & (220) & (297) & (296) &(217) \\
& & & & & & \\
$a$  (AU) & $0.049_{-.001}^{+.001}$ & $0.123_{-.003}^{+.003}$ &  $0.130_{-.003}^{+.003}$ &  $0.152_{-.003}^{+.003}$ &  $0.187_{-0.004}^{+0.004}$ &  $0.268_{-0.006}^{+0.006}$\\
& (0.0494) & (0.123) & (0.130) & (0.152) & (0.187) & (0.269)  \\
& & & & & &\\
$M \sin i$  (M$_{\earth}$) & $5.4_{-0.3}^{+0.3}$ & $4.8_{-0.5}^{+0.5}$ &  $3.1_{-0.5}^{+0.4}$ &  $2.4_{-0.4}^{+0.4}$ &  $3.4_{-0.5}^{+0.5}$ &  $5.4_{-0.6}^{+0.5}$ \\
& (5.44) & (5.01) & (3.17) & (2.62) & (3.67) & (5.34) \\
& & & & & & \\
Periastron & $4495.6_{-0.7}^{+0.8}$ & $4472_{-5}^{+5}$ &  $4468_{-4}^{+4}$ &  $4480_{-2}^{+3}$ &  $4447_{-2}^{+2}$ &  $4387_{-4}^{+5}$  \\
\ passage &  (4503.0) & (4499)& (4503) & (4480) & (4499) & (4477) \\
\ (JD - 2,450,000) & & & & & & \\
\hline
\end{tabular}
\end{minipage}
\end{table*}

\section{Discussion}
\label{sec:discussion}

We first consider whether any of the 6 signals detected is the result of stellar activity. \cite{Delfosse2012} analyzed several stellar activity diagnostics and found a high peak in one diagnostic (the full-width at half-maximum (FWHM) of the crosscorrelation
function) with a period of $\sim 105$ d, which they interpret as the rotation period of the star. In our Kepler periodgrams a 106 d period was detected starting at the 2 planet model which transitioned to a 53 d period commencing with the 5 planet model. Clearly these two periods are harmonically related. If we assume that the 106 d period is the star rotation period then depending on the configuration of surface activity (eg., spots) it would not be too surprising to detect RV variations at the second harmonic of the rotation frequency. This is a possible interpretation of the 53 d period feature in our analysis.

\begin{figure}
\includegraphics[width=85mm]{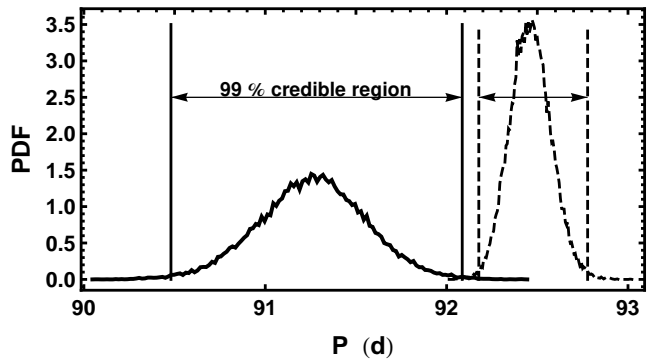}
\caption{The marginal distribution of the 91.26 d signal (solid) is compared to the marginal distribution of the 30.82 d signal (dashed) after multiplying its period scale by a factor of 3. The dashed and solid vertical lines indicate the 99\% credible regions of the two signals.}
\label{fig:6planP3P6Marg}
\end{figure}

In Sec.~\ref{sec:6HARPS}, we established that the 30.82 d period is not an alias of the 28.14 d period. Fig.~\ref{fig:6planP3P6Marg} demonstrates that the 30.82 is not a harmonic of the 91.26 d period, both of which show up in the 6 plan and 7 planet Kepler periodograms. 

\begin{figure}
\includegraphics[width=85mm]{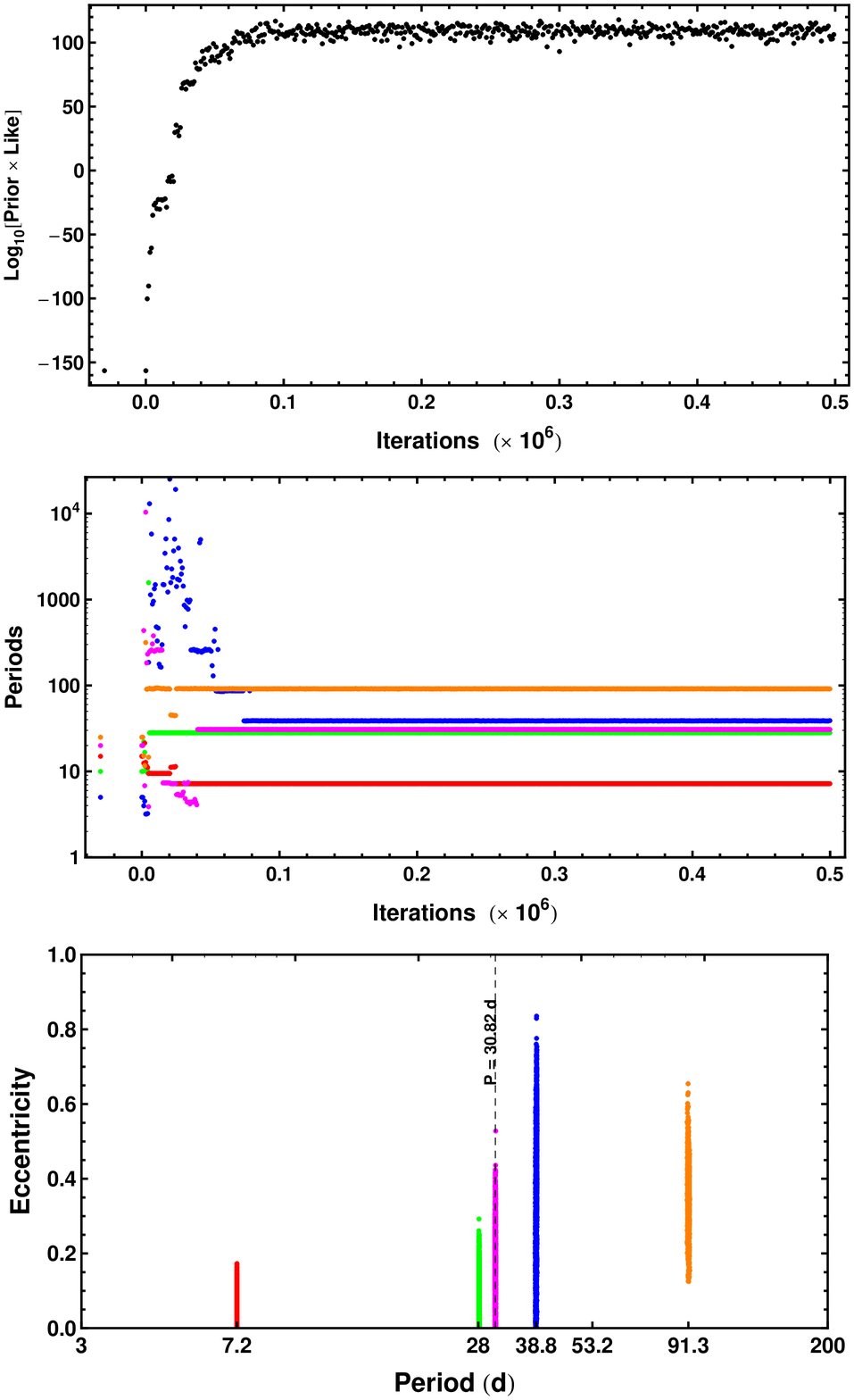}
\caption{The upper panel is a plot of the Log$_{10}$[Prior $\times$ Likelihood] versus iteration for a 5 planet Kepler periodogram of the HARPS data with 53 d signal subtracted. The middle shows the values of the five period parameters versus iteration number. The five starting periods of 5, 10, 15, 20, 25 d are shown on the left hand side of the plot at a negative iteration number. The bottom panel is a plot of the eccentricity parameter versus period for post burn-in iterations.}
\label{fig:5planMinus53dResidPiterEccP}
\end{figure}

As another test of the stability of the 6 planet model results, we subtracted one of the signals at a time (for the 91, 53, and 28 d signals) from the data and carried out a 5 planet Kepler periodogram of the modified data starting from an initial period set of 5, 10, 15, 20, and 25 d. In each case, we recovered the other 5 periods. Fig.~\ref{fig:5planMinus53dResidPiterEccP} shows the 5 planet periodogram result of the data with the 53 d signal subtracted. The upper panel is a plot of the Log$_{10}$[Prior $\times$ Likelihood] versus iteration for a 5 planet Kepler periodogram. The middle panel shows the values of the five period parameters versus iteration number. The five starting periods of 5, 10, 15, 20, 25 d are shown on the left hand side of the plot at a negative iteration number. The bottom panel is a plot of the eccentricity versus period for post burn-in iterations. The marginal distributions for the 7.2, 28.1, 30.8, 38.8, \& 91 d signals are indistinguishable from those found for the 6 planet Kepler periodogram of the original data. This example also demonstrates the ability of the FMCMC based Kepler periodogram to carry out a blind search in 28 parameters, including 5 period parameters which have a huge prior range. In fact, detection of the six signals found in Sec.~\ref{sec:6HARPS} would be a difficult feat for any analysis method that relies on a conventional periodogram of the residuals of an $n$ signal fit to estimate the starting period for the $n+1$ signal search. The ability of the FMCMC algorithm to explore a multi-modal environment has proven to be very important in this analysis.

The prospect exists that digging deeper with a more powerful statistical algorithm might also uncover new Keplerian-like spectral artifacts of the radial velocity measurement system or of the star's surface activity. Therefore, N-body simulations will be required to determine which combination of the signals are consistent with a stable planetary system. 

In the rest of this discusion we speculate on the possibility that some or all of the remaining 5 signals detected are planetary in origin. One surprise is the closeness of the 28 d and 30.8 d orbits whose semi-major axes differ by only 0.007 AU. Closest approach would occur every $1/(1/28.138-1/30.820) = 323$ d. There is precedence for systems with very close separations. One of the Kepler mission's many interesting findings is that many stars host multiple planets in surprisingly close orbits. Examples include: Kepler 36b \& c ($M = 4.3\ \& \ 7.8 $ M$_{\earth}$, \citealt{Carter2012}) with a minimum separation of 0.014 AU, Kepler 42b \& d ($M \le 2.7\ \& \ \le 0.9 $ M$_{\earth}$, \citealt{Muirhead2012}) with a minimum separation of 0.0038 AU, and KOI-55b \& c ($M \sim 0.44\ \& \sim 0.66$ M$_{\earth}$, \citealt{Charpinet2011})  with a minimum separation of 0.0016 AU. The masses of the parent star for Kepler 36, Kepler 42 and KOI-55 are 1.07, 0.13 and 0.496  M$_{\sun}$, respectively. Kepler 36b \& c both show transit timing variations consistent with a gravitational interaction between the two planets while no such timing variations were detected for the three Kepler 42 planets. 

\subsection{Multiple planets in the habitable zone?}
\label{sec:HZ}

By definition, the habitable zone around a host star is the region where liquid water can be stable on the surface of a rocky planet (\citealt{Huang1959}, \citealt{Kasting1993}).
In the current absence of observational constraints, \cite{Selsis2007} chose to
assess the habitable zone for planets with as few assumptions
as possible on their physical and chemical nature. Only two conditions were assumed although  they may derive from complex geophysical properties. Below we summarize these assumptions and some of the key concepts.

\noindent 1) The amount of superficial water must be large enough so that the surface can host liquid water for any temperature between
the temperature at the triple point of water, 273 K, and the critical temperature of water,  647 K. This condition implies that the water reservoir produces a surface pressure
higher than 220 bars when fully vaporized. With an Earth gravity, this corresponds to a 2.2 km layer of water, slightly lower than the mean depth of Earth oceans of 2.7 km. For a
gravity twice that of Earth, this pressure corresponds to half this depth. Planets with less water may still be habitable, but their HZ may be somewhat narrower because liquid water would disappear at a lower surface temperature.

\noindent 2) Atmospheric CO$_2$ must accumulate in a planet's atmosphere
whenever the mean surface temperature falls below the freezing point of water.  \cite{Walker1981} proposed that Earth's long-term climate was buffered by a negative feedback mechanism involving atmospheric CO$_2$ levels and the dependence of silicate weathering rates on climate. In the carbonate-silicate cycle, the CO$_2$ emitted by volcanoes is dissolved in rain water forming carbonic acid (H$_2$CO$_3$) which weathers silicate rocks. The products of silicate weathering, calcium (Ca$^{++}$),  bicarbonate (HCO$_3^{-}$) ions and dissolved silica (SiO$_2$), are transported by streams and rivers to the ocean. There,
organisms, such as foraminifera, diatoms and radiolarians, use the products to make shells of calcium carbonate (CaCO$_3$). Most of the shells redissolve,
but a fraction of them survive and are buried in sediments on the seafloor. The
combination of silicate weathering plus carbonate precipitation can be represented
chemically by CO$_2\ +\ $ CaSiO$_3\ \rightarrow \ $CaCO$_3\ + \ $SiO$_2$. The seabed is eventually subducted where the heat pushes the carbonate-silicate cycle reversible reaction in the opposite direction, eventually releasing CO$_2$ through volcanism. 

This cycle replenishes all the CO$_2$ in the combined atmosphere-ocean system on a
timescale of approximately half a million years. It is thus too slow to counteract human-induced global warming, but fast enough to have a dominating effect on the billion-year timescale of planetary evolution  \citep{Kasting2003}. Weathering rates increase
both because of the direct effect of temperature on chemical reaction rates and
because evaporation (and, hence, precipitation) rates increase as the surface temperature increases. If we increase the orbital radius, the surface temperature falls and the weathering decreases allowing atmospheric CO$_2$ to build up leading to a stabilization of the temperature through the greenhouse effect. This negative feedback in the carbonate-silicate cycle stabilizes the long-term surface temperature and the amount of CO$_2$ in the atmosphere of the Earth. Such an assumption implies that the planet is geologically active and continuously outgassing CO$_2$. It also implies that carbonates form in the presence of surface liquid water, which may require continental weathering. 

At a sufficiently large orbital radius the planet gets cold enough that water vapor disappears followed by the freezing of CO$_2$ which results in the permanent loss of both greenhouse gases (positive feedback). With no atmospheric CO$_2$, or with a fixed CO$_2$ level as in \cite{Hart1979}, the HZ could be $\sim 10$ times narrower. In the absence of a greenhouse gas like CO$_2$ and H$_2$O, the present Earth would be frozen.

With decreasing orbital radius, a positive feedback eventually takes over in which too much water accumulate in the atmosphere causing the temperature to increase. At a sufficiently high temperature water stops condensing. With no rain the weathering ceases and CO$_2$ builds up, further increasing the temperature. Eventually the oceans evaporate completely. Water in the high atmosphere is dissociated into hydrogen and oxygen, the hydrogen escapes to space, the oxygen combines with minerals and all the water disappears.

According to \cite{Selsis2007}, planets with masses outside the 0.5 - 10 M$_{\earth}$ range cannot host liquid surface
water. Planets under the lower end of this range have too weak a gravity to retain a sufficiently dense atmosphere, and those above the upper end accrete a massive He-H envelope. In either case, the pressure at the surface is incompatible with liquid
water. The 10 M$_{\earth}$ upper limit is somewhat fuzzy, since planets in
the 3 - 10 M$_{\earth}$ range can have very different densities.

\begin{figure}
\includegraphics[width=85mm]{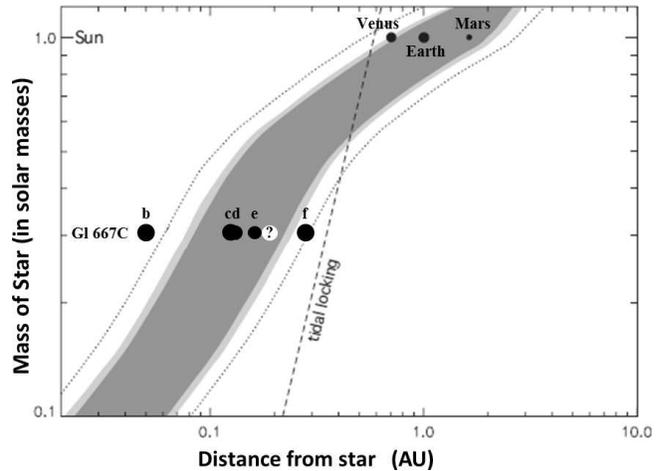}
\caption{The darker area is the orbital region that remains continuously habitable during
at least 5 Gyr as a function of the stellar mass \citep{Selsis2007}. The
light grey region gives the theoretical inner (runaway greenhouse) and
outer limits with 50\% cloudiness, with H$_2$O and CO$_2$ clouds, respectively.
The dotted boundaries correspond to the extreme theoretical limits,
found with a 100\% cloud cover. The dashed line indicates the distance
at which a 1 M$_{\earth}$ planet on a circular orbit becomes tidally locked
in less than 1 Gyr.}
\label{fig:SelsisHZ}
\end{figure}

Based on the above two major assumptions, \citep{Selsis2007} derives boundaries for the HZ as show in Fig.~\ref{fig:SelsisHZ} for a range of stellar masses. Assuming a planetary origin for the Keplerian-like signals reported here, the 28.1, 30.8, and 38.8 d signals (labeled c, d, and e) translate to orbits in the centre of the HZ. The semi-major axis of a 91.3 d planet (f) would lie just within the outermost extreme edge of the HZ, although its eccentric orbit will result in it spending the majority of its time outside the HZ. The white object in the diagram corresponds to the orbital parameters implied by a 53 d Keplerian signal. As explained above, it is likely that the star's surface activity is responsible for this signal as it is the second harmonic of what is believed to be the star's rotation period.  

\section{Conclusions}

A Bayesian re-analysis of published HARPS precision radial velocity data \cite{Delfosse2012} for Gl 667C was carried out with a multi-planet Kepler periodogram (from 0 to 7 planets) based on our fusion Markov chain Monte Carlo algorithm. In all cases the analysis included an unknown parameterized stellar jitter noise term and an unknown $slope$ parameter to deal with the linear drift resulting from the orbital motion of Gl 667C relative to the center of mass of the Gl 667ABC system. The most probable number of signals detected is 6 with a Bayesian false alarm probability of 0.012. The residuals are shown to be consistent with white noise. The Keplerian signals detected include the 7.2 and 28.1 d (planets b and c) periods reported previously plus periods of 30.8 (d), 38.8 (e), 53.2, and 91.3 (f) d. The 53.2 d period appears to correspond to the second harmonic of the stellar rotation period and is likely the result of surface activity (eg., spots). The same set of periods were also detected in a subset of the data consisting of the first 143 data points. 

The existence of the additonal Keplerian-like signals suggest the possibilty of further planets, two of which (d and e) could join Gl 667Cc in the central region of the habitable zone. $M \sin i$ values corresponding to signals b, c, d, e, and f are $\sim$ 5.4, 4.8, 3.1, 2.4, and 5.4 M$_{\earth}$, respectively. It is also possible that digging deeper with a more powerful statistical algorithm might have uncovered new spectral artifacts of the radial velocity measurement system or of the star's surface activity. N-body simulations are required to determine which of these signals are consistent with a stable planetary system. 

\section*{Acknowledgments}

The author would like to thank Wolfram Research for providing a complementary license to run gridMathematica to facilitate parallel processing.

\end{document}